\documentclass[preprint,12pt]{aastex}

\shorttitle{Rotation or Anisotropy}

\shortauthors{Rothberg \& Joseph}

\begin{document}
\title{A Survey of Merger Remnants III: Are Merger Remnants Supported by Rotation or Anisotropy?
\footnote{Some of the data
presented herein were obtained at the W.M. Keck Observatory, which is operated as a scientific
partnership among the California Institute of Technology, the University of California and the
National Aeronautics and Space Administration. The Observatory was made possible by the generous
financial support of the W.M. Keck Foundation.}}

\author{B. Rothberg}
\affil{Space Telescope Science Institute, 3700 San Martin Drive Baltimore, MD 21218}
\email{rothberg@stsci.edu}

\author{R. D. Joseph}
\affil{Institute for Astronomy, 2680 Woodlawn Drive, Honolulu, HI 96822}

\begin{abstract}
\indent  A growing body of observational evidence suggests that the luminosity, photometric shape and amount of 
rotational or anisotropical support in elliptical galaxies may provide vital clues to how they formed.  Elliptical galaxies 
appear to fall into two distinct categories based on these parameters:  bright, boxy shaped, with little or no rotation, 
and less luminous, disky shaped with significant rotation.  One viable formation scenario is 
the ``Merger hypothesis,'' in which two disk galaxies merge to form a new elliptical galaxy.   
A comparison of the luminosity, photometric shape, and amount of rotation in advanced merger remnants may shed 
more light on the possible formation scenarios of elliptical galaxies.  Yet, little observational data exists
for such merger remnants.  This paper is the third in a series investigating the photometric and kinematic 
properties of a sample of 51 {\it optically} selected advanced merger remnants.  Presented here are
{\it K}-band isophotal shapes and spatially resolved kinematics for a sub-sample of 37 merger remnants.
The results show that $\sim$ 11$\%$ of the sample are boxy, and anisotropically supported
while $\sim$ 47$\%$ are disky, and rotationally supported. The remainder of the sample
show variations among expected correlations between shape and rotation.  
This may suggest that the isophotal shapes are still ``in flux.'' There does appear to be a {\it lower} 
limit to the amount of anisotropy observed in the merger remnants.  This {\it may} provide an observational 
diagnostic for discriminating among formation scenarios in elliptical galaxies.
\end{abstract}


\section{Introduction}
\indent Observations of elliptical galaxies reveal that they are far more complicated than first thought.
Rather than being homogeneous ellipsoids flattened by rotation, photometry has shown that many
elliptical galaxies exhibit isophotal deviations from perfect ellipses
(e.g. \markcite{1978MNRAS.182..797C,1985MNRAS.216..429L}{Carter} (1978); {Lauer} (1985)).  Kinematic measurements indicate that
some ellipticals show smaller than expected rotation \markcite{1977ApJ...218L..43I}({Illingworth} 1977), while 
others appear to show significant rotation \markcite{1983ApJ...266...41D}({Davies} {et~al.} 1983).  Theoretical modeling
has suggested that those which show smaller than expected rotation may be either triaxial in shape or flattened
by anisotropy \markcite{1976MNRAS.177...19B, 1978MNRAS.183..501B}({Binney} 1976, 1978).  Isophotal shapes also appear to correlate
with luminosity and rotation \markcite{1988A&AS...74..385B,1989A&A...217...35B}({Bender}, {Doebereiner}, \&  {Moellenhoff} 1988; {Bender} {et~al.} 1989), (hereafter BDM88 and B89). 
Follow-up studies have revealed that other properties such as X-ray and radio 
luminosities are correlated with boxy ellipticals, but not disky ellipticals (B89). \\ 
\indent  These observational results have formed a paradigm for classifying
elliptical galaxies as either luminous, boxy-shaped, with little or no rotation (anisotropy), 
or faint, disky-shaped, with significant rotation. This now well-established dichotomy between ``boxy'' and ``disky'' 
ellipticals has fueled speculation that these properties are linked to different formation mechanisms.  
\markcite{1996ApJ...464L.119K}{Kormendy} \& {Bender} (1996) and \markcite{1989ARA&A..27..235K}{Kormendy} \& {Djorgovski} (1989) suggest a new morphological classification scheme in which 
ellipticals are categorized by their isophotal shapes.   They argue that photometric shapes are a reliable measure 
of velocity anisotropy, which in turn is directly connected to how elliptical galaxies form.  Furthermore, they suggest that 
disky isophotal shapes may be the result of gaseous dissipation which occurred during their formation.  
The origin of the dichotomy between boxy and disky ellipticals leads to the question of what role mergers may play 
in the formation of elliptical galaxies.\\
\indent In the context of the Toomre Merger Hypothesis \markcite{1972ApJ...178..623T,1977egsp.conf..401T}({Toomre} \& {Toomre} 1972; {Toomre} 1977) it is clear how disk-disk
mergers can produce elliptical galaxies.  However, the importance and relative contributions of gaseous dissipation, 
isophotal shapes and kinematics is an area which has, and still continues, to generate some controversy.  Gaseous dissipation 
has often 
been linked with mergers. For example, \markcite{1994ApJ...437L..47M,2000MNRAS.312..859S}{Mihos} \& {Hernquist} (1994); {Springel} (2000) 
included dissipation in their numerical simulations, which produced simulated merger remnants with ``excess light'' 
in their surface brightness profiles.  This feature was later observed in one-third of the merger remnants in  
\markcite{2004AJ....128.2098R}{Rothberg} \& {Joseph} (2004), hereafter Paper I.  Both numerical simulations \markcite{1996ApJ...471..115B,2002MNRAS.333..481B}({Barnes} \& {Hernquist} 1996; {Barnes} 2002) 
and observations \markcite{1992ApJ...396..510W,1993AJ....106.1354W}({Wang}, {Schweizer}, \&  {Scoville} 1992; {Whitmore} {et~al.} 1993) show that dissipation can lead to the formation of 
central gaseous disks which could later form stellar disks.  The effects of gaseous dissipation were observationally 
inferred from the results in \markcite{2006AJ....131..185R}{Rothberg} \& {Joseph} (2006), hereafter Paper II.  The kinematic and photometric data showed
that the central phase-space densities of the merger remnants were consistent with those of
elliptical galaxies.  One way to increase the phase-space density of spiral disks to match the levels in elliptical
galaxies is a starburst induced by the dissipative collapse of gas 
(e.g. \markcite{1983MNRAS.205.1009N,1985MNRAS.214...87J,1987nngp.proc...18S,1991ApJ...370L..65B,
1992ApJ...390L..53K,1994ApJ...437L..47M,1996ApJ...464..641M}{Negroponte} \& {White} (1983); {Joseph} \& {Wright} (1985); {Schweizer} (1987); {Barnes} \& {Hernquist} (1991); {Kormendy} \& {Sanders} (1992); {Mihos} \& {Hernquist} (1994, 1996)).  \\
\indent However, questions remain as to whether isophotal shapes are linked to specific formation
scenarios.  It has been argued that boxy ellipticals are also produced by mergers 
\markcite{1985MNRAS.214..449B,1993MNRAS.261..379G}({Binney} \& {Petrou} 1985; {Governato}, {Reduzzi}, \&  {Rampazzo} 1993).  \markcite{1990ApJ...364L..33S}{Schweizer} {et~al.} (1990) even include 
boxiness in their definition of the fine structure index $\Sigma$, which they use to identify ellipticals 
with possible merger origins.  Dissipationless (gas-free) merger models by \markcite{1999ApJ...523L.133N}{Naab}, {Burkert}, \&  {Hernquist} (1999) and
\markcite{2003ApJ...597..893N}{Naab} \& {Burkert} (2003), hereafter NB03, show that equal-mass and some 2:1 mass ratio
merger remnants can produce boxy shaped remnants which are anisotropically supported. 
A similar result was found for earlier numerical simulations by \markcite{1994ApJ...427..165H}{Heyl}, {Hernquist}, \&  {Spergel} (1994).  They concluded that
the formation of elliptical galaxies via dissipationless merging does not lead preferentially to boxy or disky
isophotes.  All of these simulations found that viewing angle contributed significantly to whether a simulated
merger remnant could be classified as boxy or disky.  However, it is not clear
what effect the presence of gas and star-formation may have on isophotal shapes or kinematics.  There is some evidence 
that gas may suppress box orbits in mergers \markcite{1996ApJ...471..115B}({Barnes} \& {Hernquist} 1996).  New results from  \markcite{2006MNRAS..inprep}{Naab} {et~al.} (2006)
(hereafter N06), which include a non-star-forming gaseous component, indicate that 1:1 mass ratio merger remnants produce
a wider range of kinematic properties, but seem more likely to have disky isophotal shapes.\\
\indent The sample of advanced merger remnants presented in Papers I and II are ideal for testing the
complicated and seemingly contradictory information offered by both numerical simulations and observations of
elliptical galaxies.  The {\it K}-band photometry from Paper I, in conjunction with the spectroscopic data (centered 
on the Ca triplet lines at $\lambda$ $\sim$ 8500 {\AA}) from Paper II, will be used to measure the degree of rotational or 
anisotropic support in the merger remnants and compare that with their isophotal shapes and luminosity. 
The {\it K}-band is best suited for measuring the stellar component of mergers
for two reasons.  First, dust extinction is approximately one-tenth that at the {\it V}-band.  Second, the
blackbody emission of older, late-type stars, which trace the majority of the stellar mass in both elliptical
and spiral galaxies, peaks in the near-infrared beyond 1 $\micron$ \markcite{1981IAUS...96..297A}({Aaronson} 1981).  If mergers
are undergoing star-formation, then the optical light is likely to be dominated by young stars 
and/or be heavily extincted.  The Ca triplet
lines were selected to measure stellar velocities and dispersions because
they are strong, well-separated, and narrow \markcite{1984ApJ...286...97D}({Dressler} 1984).  Absorption lines at shorter wavelengths
are more likely to be contaminated by strong nearby emission lines from galaxies undergoing star-formation or with 
young stellar populations, as well as more sensitive to template mismatch.  Finally, the redward Ca triplet lines
are somewhat less affected by extinction than absorption lines at shorter wavelengths.\\
\indent The results will also be compared with real elliptical galaxies, primarily from BDM88 and B89, 
and numerical simulations of dissipationless merger remnants from NB03 along with simulated merger remnants with a 
gaseous component from N06.  The goal is to test whether mergers are capable of forming boxy or disky (or both) elliptical galaxies.
However, an important caveat to bear in mind is that the simulated merger remnants from NB03 are the result of
purely stellar interactions and merging, while the results from Paper I and II indicate that the sample of observed merger remnants  
appear to be consistent with having undergone some form of gaseous dissipation and star-formation.  Although N06 
does include a gaseous component in the simulated merger remnants, the simulations do not include star-formation.  
The presence of star-formation could lead to differences between the observed and simulated merger remnants.
All data and calculations presented in this paper assume a value of {\it H}$_{\circ}$ = 75 km s$^{-1}$ Mpc$^{-1}$.

\section{Samples}
\subsection{Merger Sample}
\indent The sample of merger remnants presented here is tailored to be reasonably large and to include only
objects in the advanced stages of merging.  The sample selection was based {\it solely} on optical morphology 
because, by definition, merger identification is a visual selection criterion.  Selections based on other criteria, 
such as far-infrared luminosities, may introduce a bias into the sample towards objects undergoing star-formation. 
The 51 objects in the sample were chosen primarily from the Atlas of 
Peculiar Galaxies \markcite{1966apga.book.....A}({Arp} 1966), A Catalogue of Southern Peculiar Galaxies 
\markcite{1987cspg.book.....A}({Arp} \& {Madore} 1987), the Atlas and Catalog of Interacting Galaxies \markcite{1959VV....C......0V}({Vorontsov-Velyaminov} 1959),
and the Uppsala General Catalogue \markcite{1973ugcg.book.....N}({Nilson} 1973). 
The full sample was selected based strictly on optical morphology and according to the following
criteria:\\
\indent 1)  Tidal tails, loops, and shells, which are induced by strong gravitational
            interaction.\\
\indent 2)  A single nucleus, which, based on numerical studies,
        marks the completion of the merger.  This criterion is important because it marks the point
        at which the merger should begin to exhibit properties in common with elliptical galaxies.\\
\indent 3)  The absence of nearby companions which may induce the presence of tidal
        tails and make the object appear to be in a more advanced stage of merging.\\
\indent 4)  The merger remnants must be observable from Mauna Kea.  This limited the survey to objects
        with declinations $\geq$ -50$\degr$.\\
\indent Spectroscopic observations were made for a sub-sample of 38 objects.  The sub-sample was not 
selected based on any particular criteria.  Initially, spectroscopic observations were planned for all 
51 objects in the sample, yet due to limitations of time, not all of the objects could be observed.  
While spectroscopic data were obtained for 38 objects, only 37 of these had sufficient signal-to-noise (S/N) to 
extract kinematic data spatially along the slit.\\
\indent The merger remnants presented in this paper are listed in Table 1 which includes names, right ascension, declination and
derived heliocentric recessional velocities.  Since most of the objects have multiple designations, all subsequent references
to sample galaxies within the paper, tables and figures will first use the NGC designation if available,  
followed by the Arp or Arp-Madore (AM), UGC, VV and lastly the IC designation if no other designation 
is available.  Unless otherwise noted, the galaxies are listed in order of Right Ascension in 
tables and figures.  \\
\indent Within the merger sample there are three sub-samples, ``shell ellipticals,'' 
Ultra-luminous and luminous infrared galaxies (ULIRG/LIRGs), and ''normal merger remnants,'' which 
are defined simply as those galaxies which are neither ULIRG/LIRGs nor shell ellipticals.  
These distinctions are noted in Table 1.  However, for the purposes of this paper, no
distinction is made in the analysis among the sub-samples.
\clearpage
{
\begin{deluxetable}{llcccc}
\tabletypesize{\normalsize}
\setlength{\tabcolsep}{0.1in}
\tablewidth{0pt}
\tablenum{1}
\pagestyle{empty}
\tablecaption{Merger Kinematic Sub-Sample}
\tablecolumns{6}
\tablehead{
\colhead{Merger Names} &
\colhead{Other Names} &
\colhead{R.A.} &
\colhead{Dec.} &
\colhead{{\it V$_{\odot}$}\tablenotemark{a}} &
\colhead{notes}\\
\colhead{} &
\colhead{} &
\colhead{(J2000)}&
\colhead{(J2000)}&
\colhead{(km s$^{-1}$)} &
\colhead{}
}
\startdata
UGC 6       &VV 806                     &00$^{h}$ 03$^{m}$ 09$^{s}$ &21$^{\circ}$ 57$^{'}$ 37$^{''}$  &6579  &LIRG\\
NGC 34      &VV 850                     &00$^{h}$ 11$^{m}$ 06$^{s}$ &-12$^{\circ}$ 06$^{'}$ 26$^{''}$ &5881  &LIRG\\
NGC 455     &Arp 164, UGC 815           &01$^{h}$ 15$^{m}$ 57$^{s}$ &05$^{\circ}$ 10$^{'}$ 43$^{''}$  &5827  &    \\
NGC 1210    &AM 0304-255                &03$^{h}$ 06$^{m}$ 45$^{s}$ &-25$^{\circ}$ 42$^{'}$ 59$^{''}$ &3878  &Shell\\
NGC 1614    &Arp 186                    &04$^{h}$ 33$^{m}$ 59$^{s}$ &-08$^{\circ}$ 34$^{'}$ 44$^{''}$ &4769  &LIRG \\
AM 0612-373 &\nodata                    &06$^{h}$ 13$^{m}$ 47$^{s}$ &-37$^{\circ}$ 40$^{'}$ 37$^{''}$ &9721  &     \\
NGC 2418    &Arp 165, UGC 3931          &07$^{h}$ 36$^{m}$ 37$^{s}$ &17$^{\circ}$ 53$^{'}$ 02$^{''}$  &5037  &     \\
NGC 2623    &Arp 243, UGC 4509, VV 79   &08$^{h}$ 38$^{m}$ 24$^{s}$ &25$^{\circ}$ 45$^{'}$ 17$^{''}$  &5549  &LIRG \\
UGC 4635    &\nodata                    &08$^{h}$ 51$^{m}$ 54$^{s}$ &40$^{\circ}$ 50$^{'}$ 09$^{''}$  &8722  &     \\
NGC 2655    &Arp 225, UGC 4637          &08$^{h}$ 55$^{m}$ 37$^{s}$ &78$^{\circ}$ 13$^{'}$ 23$^{''}$  &1400  &Shell\\
NGC 2782    &Arp 215, UGC 4862          &09$^{h}$ 14$^{m}$ 05$^{s}$ &40$^{\circ}$ 06$^{'}$ 49$^{''}$  &2543  &LIRG \\
NGC 2914    &Arp 137, UGC 5096          &09$^{h}$ 34$^{m}$ 02$^{s}$ &10$^{\circ}$ 06$^{'}$ 31$^{''}$  &3159  &     \\
UGC 5101    &\nodata                    &09$^{h}$ 35$^{m}$ 51$^{s}$ &61$^{\circ}$ 21$^{'}$ 11$^{''}$  &11802 &ULIRG\\
NGC 3256    &AM 1025-433, VV 65         &10$^{h}$ 27$^{m}$ 51$^{s}$ &-43$^{\circ}$ 54$^{'}$ 14$^{''}$ &2795  &LIRG \\
Arp 156     &UGC 5814                   &10$^{h}$ 42$^{m}$ 38$^{s}$ &77$^{\circ}$ 29$^{'}$ 41$^{''}$  &10738 &     \\
NGC 3597    &AM 1112-232                &11$^{h}$ 14$^{m}$ 41$^{s}$ &-23$^{\circ}$ 43$^{'}$ 39$^{''}$ &3500  &     \\
NGC 3921    &Arp 224, UGC 6823, VV 31   &11$^{h}$ 51$^{m}$ 06$^{s}$ &55$^{\circ}$ 04$^{'}$ 43$^{''}$  &5896  &     \\
NGC 4004    &UGC 6950, VV 230           &11$^{h}$ 58$^{m}$ 05$^{s}$ &27$^{\circ}$ 52$^{'}$ 44$^{''}$  &3368  &     \\
NGC 4194    &Arp 160, UGC 7241, VV 261  &12$^{h}$ 14$^{m}$ 09$^{s}$ &54$^{\circ}$ 31$^{'}$ 36$^{''}$  &2501  &LIRG \\
NGC 4441    &UGC 7572                   &12$^{h}$ 27$^{m}$ 20$^{s}$ &64$^{\circ}$ 48$^{'}$ 06$^{''}$  &2722  &     \\
AM 1255-430 &\nodata                    &12$^{h}$ 58$^{m}$ 08$^{s}$ &-43$^{\circ}$ 19$^{'}$ 47$^{''}$ &8984  &     \\
NGC 5018    &UGCA 335                   &13$^{h}$ 13$^{m}$ 00$^{s}$ &-19$^{\circ}$ 31$^{'}$ 05$^{''}$ &2816  &Shell\\
Arp 193     &UGC 8387, VV 821, IC 883   &13$^{h}$ 20$^{m}$ 35$^{s}$ &34$^{\circ}$ 08$^{'}$ 22$^{''}$  &6985  &LIRG \\
AM 1419-263 &\nodata                    &14$^{h}$ 22$^{m}$ 06$^{s}$ &-26$^{\circ}$ 51$^{'}$ 27$^{''}$ &6758  &     \\
UGC 9829    &VV 847                     &15$^{h}$ 23$^{m}$ 01$^{s}$ &-01$^{\circ}$ 20$^{'}$ 50$^{''}$ &8465  &     \\
NGC 6052    &Arp 209, UGC 10182, VV 86  &16$^{h}$ 05$^{m}$ 12$^{s}$ &20$^{\circ}$ 32$^{'}$ 32$^{''}$  &4739  &     \\
UGC 10607   &VV 852, IC 4630            &16$^{h}$ 55$^{m}$ 09$^{s}$ &26$^{\circ}$ 39$^{'}$ 46$^{''}$  &10376 &     \\
UGC 10675   &VV 805                     &17$^{h}$ 03$^{m}$ 15$^{s}$ &31$^{\circ}$ 27$^{'}$ 29$^{''}$  &10179 &     \\
AM 2038-382 &\nodata                    &20$^{h}$ 41$^{m}$ 13$^{s}$ &-38$^{\circ}$ 11$^{'}$ 36$^{''}$ &6092  &     \\
AM 2055-425 &\nodata                    &20$^{h}$ 58$^{m}$ 26$^{s}$ &-42$^{\circ}$ 39$^{'}$ 00$^{''}$ &12890 &LIRG \\
NGC 7135    &AM 2146-350, IC 5136       &21$^{h}$ 49$^{m}$ 46$^{s}$ &-34$^{\circ}$ 52$^{'}$ 35$^{''}$ &2644  &     \\
UGC 11905   &\nodata                    &22$^{h}$ 05$^{m}$ 54$^{s}$ &20$^{\circ}$ 38$^{'}$ 22$^{''}$  &7456  &     \\
NGC 7252    &Arp 226, AM 2217-245       &22$^{h}$ 20$^{m}$ 44$^{s}$ &-24$^{\circ}$ 40$^{'}$ 41$^{''}$ &4792  &     \\
AM 2246-490 &\nodata                    &22$^{h}$ 49$^{m}$ 39$^{s}$ &-48$^{\circ}$ 50$^{'}$ 58$^{''}$ &12901 &ULIRG\\
IC 5298     &\nodata                    &23$^{h}$ 16$^{m}$ 00$^{s}$ &25$^{\circ}$ 33$^{'}$ 24$^{''}$  &8221  &LIRG \\
NGC 7585    &Arp 223                    &23$^{h}$ 18$^{m}$ 01$^{s}$ &-04$^{\circ}$ 39$^{'}$ 01$^{''}$ &3539  &Shell\\
NGC 7727    &Arp 222, VV 67             &23$^{h}$ 39$^{m}$ 53$^{s}$ &-12$^{\circ}$ 17$^{'}$ 35$^{''}$ &1868  &     \\
\enddata
\tablecomments{(a) $\sigma$$_{\circ}$ (1.53 kpc aperture) measured from Ca triplet lines}
\end{deluxetable}

}
\clearpage
\subsection{Comparison Samples of Elliptical Galaxies and Simulated Merger Remnants}
\indent The same sets of kinematic and photometric properties presented in this paper for the observed merger remnants
were extracted from a sample of 141 elliptical galaxies from \markcite{1988A&AS...74..385B,1992ApJ...399..462B, 1994MNRAS.269..785B}{Bender} {et~al.} (1988); {Bender}, {Burstein}, \&  {Faber} (1992); {Bender}, {Saglia}, \&  {Gerhard} (1994), 
and additional unpublished data from Bender.  This is the same data set used by \markcite{1999ApJ...523L.133N}{Naab} {et~al.} (1999), NB03 and N06 to 
compare with their numerical simulations.  However, the data set did not include {\it K}-band photometry.  Supplemental {\it K}-band 
data were obtained from the 2MASS archive, including the large galaxy atlas \markcite{2003AJ....125..525J}({Jarrett} {et~al.} 2003).  A final sample of 84 
elliptical galaxies, with measured values for all of the same parameters derived for the merger remnants, was
compiled from the above data sets. \\
\indent It is also important to briefly discuss the simulated merger remnants  which
are used as a comparison with the observed merger remnants presented in this paper. The photometric and kinematic
parameters of the simulated remnants from NB03 were measured at a stage beyond which the bulge components of the progenitors have
coalesced.  Specifically, the simulated remnants merge at {\it t} = 80, where 1 time unit is $\sim$ 1.3$\times$10$^{7}$ yrs.
The simulations are allowed to evolve for 10 dynamical times ($\sim$ 10$^{8}$ yrs or 70 time units).  The evolution is then followed
from {\it t} = 150-200 to assure that no further evolution occurs.  In the central regions, the dynamical times are shorter
than 10$^{8}$ years.  The results of the simulations show little or no evolution after 150-200 time steps.  
The simulations are scale-free, therefore, if one assumes a smaller mass for the progenitors, 
the time units will scale appropriately.  \\
\indent The simulated merger remnants from N06 include a gas component. The remnants 
from NB03 were re-simulated with an additional gas component in the progenitor disk.  Only 1:1 and 3:1 mass ratio
simulations are presented here.  10$\%$ of the stellar disk
in each progenitor was replaced by gas and the two progenitor disks were then merged in a fashion similar to NB03.  
The gas was represented by smooth particle hydrodynamics assuming an isothermal equation of state.  This implies that additional heat 
created in shocks, adiabatic compression and feedback processes is radiated away immediately.  No star-formation is
included in the simulations.  The merger remnants were allowed to settle into dynamical equilibrium for approximately 
30 dynamical time-scales after the merger was complete, before their properties were measured.\\
\indent The observed merger remnants have all been selected so that the sample only includes objects with a
detectable single nucleus
at {\it K}-band.  Taking into account the effects of resolution and seeing, even the most distant merger remnant in the sample
can be resolved down to a scale of $\sim$ 670 pc.  Therefore, even if two nuclei are present but cannot be observationally separated,
the {\it kinematic} properties should still be fairly close to their final values.  Other numerical simulations have shown
that kinematic properties (i.e. velocity dispersion) settle to a constant value when the nuclei of the progenitors
are at distances $<$ 1 kpc \markcite{1999Ap&SS.266..195M}({Mihos} 1999). The observed merger remnants and the simulated merger remnants have
ages which are consistent with each other.  Regardless of how far along the merging process the observed remnants are,
their kinematic properties are directly comparable to the simulated remnants because both samples have been 
``observed'' after a stage in which these properties have been ``frozen'' into place.  However, the photometric properties,
such as ellipticity and isophotal shape may still be different.  The results from Paper I indicate that while some of the 
remnants are devoid of central structure, not all appear to have fully phase mixed.  Therefore, these properties
may still be ``in flux'' as compared with the simulated remnants.  

\section{Observations}
\subsection{{\it K}-band imaging}
\indent Near-infrared images were obtained using the Quick Infrared Camera (QUIRC) 
1024 $\times$ 1024 pixel HgCdTe infrared array \markcite{1996NewA....1..177H}({Hodapp} {et~al.} 1996) at the f/10 
focus on the University of Hawaii 2.2 meter telescope. 
The field of view of the QUIRC array is 193{$\arcsec$} x 193{$\arcsec$} with a plate scale 
of 0.189{$\arcsec$} pixel$^{-1}$.  Details of how the observations were conducted
can be found in Papers I and II.  \\
\indent The median seeing for the observations was 0{$\arcsec$}.8.
The {\it K} filter used in this survey for all but two objects conforms to the new 
Mauna Kea Infrared Filter Set \markcite{2002PASP..114..180T}({Tokunaga}, {Simons}, \&  {Vacca} 2002).  One object, UGC 6 was 
observed with the {\it K'} filter \markcite{1992AJ....103..332W}({Wainscoat} \& {Cowie} 1992) and has been converted to {\it K} 
using their conversion equation. Another object, IC 5298, was observed with an older {\it K} 
filter with similar properties to the Mauna Kea {\it K} filter.  No conversion was made 
and it is assumed that {\it K$_{old}$} $\simeq$ {\it K$_{MaunaKea}$}.  Table 2 is the observation log for both
the photometric and spectroscopic observations.  Column (2) of Table 2 lists the total
integration time for each object.
\clearpage
{
\begin{deluxetable}{lccc}
\tabletypesize{\normalsize}
\setlength{\tabcolsep}{0.05in}
\tablewidth{0pt}
\tablenum{2}
\pagestyle{empty}
\tablecaption{Observation Log}
\tablecolumns{4}
\tablehead{
\colhead{Merger Name} &
\colhead{QUIRC Integration Time} &
\colhead{ESI Integration Time} &
\colhead{Slit P.A.}  \\
\colhead{} &
\colhead{(sec)}&
\colhead{(sec)} &
\colhead{(degrees)}
}
\startdata
UGC 6       &3210 &540  &90.0  \\
NGC 34      &3300 &1200 &-41.0 \\
NGC 455     &2250 &1800 &-30.0 \\
NGC 1210    &2820 &300  &-41.0 \\
NGC 1614    &2730 &1800 &32.9  \\
AM 0612-373 &2160 &1800 &40.0  \\
NGC 2418    &3600 &1800 &30.8  \\
NGC 2623    &3390 &1800 &64.1  \\
UGC 4635    &3480 &1800 &49.8  \\
NGC 2655    &1800 &1800 &83.8  \\
NGC 2782    &3000 &1800 &90.0  \\
NGC 2914    &1800 &1800 &20.5  \\
UGC 5101    &2700 &1800 &83.0  \\
NGC 3256    &2925 &629  &0.0   \\
Arp 156     &2400 &3600 &-61.8 \\
NGC 3597    &3300 &1800 &76.7  \\
NGC 3921    &3645 &1800 &29.5  \\
NGC 4004    &1560 &1800 &-12.0 \\
NGC 4194    &3600 &1800 &-20.0 \\
NGC 4441    &1440 &1800 &2.0   \\
AM 1255-430 &3420 &2700 &-77.2 \\
NGC 5018    &2025 &1140 &90.0  \\
Arp 193     &2625 &3600 &-39.3 \\
AM 1419-263 &3600 &1800 &69.0  \\
UGC 9829    &3600 &1800 &-15.0 \\
NGC 6052    &3600 &1800 &71.5  \\
UGC 10607   &3360 &1800 &0.0   \\
UGC 10675   &3600 &1800 &90.0  \\
AM 2038-382 &1920 &1200 &-45.0 \\
AM 2055-425 &2880 &1200 &-35.0 \\
NGC 7135    &2520 &1800 &0.0   \\
UGC 11905   &3000 &1200 &49.5  \\
NGC 7252    &3360 &1800 &-60.0 \\
AM 2246-490 &2520 &1200 &-5.0  \\
IC 5298     &3240 &900  &29.7  \\
NGC 7585    &2040 &900  &-70.0 \\
NGC 7727    &3240 &900  &90.0  \\
\enddata
\end{deluxetable}

}
\clearpage
\subsection{Optical Spectroscopy}
\indent The optical spectroscopic observations were obtained with the Echellette Spectrograph and 
Imager \markcite{2002PASP..114..851S}({Sheinis} {et~al.} 2002) at the W. M. Keck-II 10-meter telescope. 
The spectrograph covers the wavelength range from 3927-11068 {\AA}.  The observations were 
made in the echelle mode using the 0{$\arcsec$}.5 $\times$ 20{$\arcsec$} slit, which
corresponds to a resolution of  {\it R} $\simeq$ 36.2 km s$^{-1}$ or {\it R} $\simeq$ 8200.  
The slit of the spectrograph was rotated to match the Position Angle (P.A.) of the
major axis of each galaxy.   The P.A. of the major axis for each galaxy was measured from the
{\it K}-band photometric data. The total integration time and the P.A. of the major 
axis are listed in columns (3) and (4) of Table 2 respectively for each galaxy.  In addition to spectrophotometric 
standards, giant stars covering the spectral range from G0III to M3III, and two super-giants of class M1Iab and M2Iab 
were observed with the same instrumental setup for use as template stars for the kinematic analysis.
Paper II describes in more detail how the observations were conducted.  The full list of template
stars observed with ESI can be found in Table 3 of Paper II.

\section{Data Reduction and Analysis}
\subsection{{\it K}-band imaging}
\indent The data set was reduced using IRAF. Structural photometric parameters were extracted from elliptical isophotes using the 
ELLIPSE task in the STSDAS package.  The radius of each isophote was taken to be the measured
seeing for that galaxy (i.e. the radius of the first isophote is the measured seeing).
Paper I describes in detail the method used to extract the {\it K}-band photometric parameters.  
For the present paper only the ellipticity ($\epsilon$) and
the a$_{4}$/a parameter were used.  The latter parameter is the amplitude of the cos 4$\theta$ term, 
which measures the deviations of the isophote from a perfect ellipse.  These two parameters were measured independent
of any fits to the surface brightness profile of a galaxy.  The ellipticity and a$_{4}$/a parameters 
used in this paper were measured relative to the spatial scales
of the kinematic observations.  The values $\bar{\epsilon}$ and $\bar{a_{4}/a}$
were taken to be the average values {\it within the radius covered by the kinematic measurements}.  These are 
the same definitions used in earlier kinematic studies of elliptical galaxies.  Table 3 lists the measured
{\it K}-band photometric parameters for each merger remnant along with derived spectroscopic parameters.
The photometric parameters $\bar{\epsilon}$ and $\bar{a_{4}/a}$
are listed in columns (2) and (3) in Table 3 respectively.\\
\indent The absolute magnitudes, {\it M}$_{K}$ were obtained by
summing up the flux in circular apertures from the center to the edge of the array, after masking out foreground stars.  
This method makes no assumptions about the profile shape of the galaxy.  
No corrections to the {\it K}-band data have been made for galactic extinction. 
Since the reddening at {\it K} is very small, any such corrections, even for objects at low galactic
latitude would be negligible.  All distance dependent values listed in Table 3 assume a value of 
{\it H}$_{\circ}$ = 75 km s$^{-1}$ Mpc$^{-1}$ and are derived using only the Heliocentric recessional
velocities listed in Table 1.  No additional corrections have been made.  The values of
{\it M}$_{K}$ are listed in column (4) of Table 3.

\subsection{Optical Spectroscopy}
\indent  Only the 7th order of the echellete spectra, containing the Ca triplet absorption line at 
$\lambda$ $\simeq$ 8500 {\AA}, was analyzed for the data presented in this paper. The details of the data reduction
are presented in Paper II.  The only difference in the method of data reduction is that multiple one-dimensional 
spectra were extracted along the spatial axis.  The aperture diameter of each
extracted spectrum is equivalent to the measured seeing.  The seeing was determined by taking the average FWHM
of several stars in the acquisition camera.  In cases where no acquisition camera images were saved, the FWHM
along the spatial axis of the spectro-photometric standards or template stars closest in time to the galaxy observation
were used.  \\
\indent The kinematic data were extracted from the spectra using the same method described in Paper II.
A single stellar template was used in the fitting for each spectra extracted spatially along the slit.  
All of the stellar templates noted in Paper II were tested with each galaxy.
The best fitting template for each galaxy was chosen based on the reduced chi-square and 
rms of the fit. The best-fit template star for each galaxy is listed in the last column of Table 3.\\
\indent The errors shown in Table 3 are not absolute, and are provided more as reasonable estimates.  The error
analysis was conducted by testing various limits on the fitting process.  First, Monte Carlo simulations were
conducted to test the fitting program.  The testing was based on 100 realizations of a template star convolved
with a Gauss-Hermite polynomial of known properties with random noise added.  This altered template star was used
as a ``test galaxy,'' to determine whether the fitting program could recover the input parameters.
Next, a second template star of identical stellar type was used to recover the input parameters of 
the ``test galaxy.''  The spread in errors from the Monte Carlo simulations were found to be nearly 
the same as the fitting errors determined by the program.  Finally, template mis-match was tested 
by investigating the spread in the derived parameters from using different template stars.  
Only template stars which produced fits within 2$\times$$\chi$$_{\nu}$$^{2}$ of  the best-fitting 
template were used to test each galaxy.  The largest errors in the fitting process were produced by
template mismatch. The errors listed in Table 3 for the kinematic parameters are the standard deviations
of the derived parameters for the range of template stars used to test the mismatch.\\
\indent The central velocity dispersions ($\sigma$$_{\circ}$) listed in column (5) of Table 3 were taken from Paper II.  
They were measured within an aperture diameter equivalent to 1.53 kpc 
(assuming {\it H}$_{\circ}$ = 75 km s$^{-1}$ Mpc$^{-1}$ and using the heliocentric velocities measured in Table 1 
for each object.  This is equivalent to a $3\farcs.4$ aperture 
at the distance of Coma.  This aperture size was selected by \markcite{1995MNRAS.273.1097J,1997MNRAS.291..461S}{Jorgensen}, {Franx}, \&  {Kjaergaard} (1995); {Smith} {et~al.} (1997) 
to bring spectroscopic measurements of $\sigma$$_{\circ}$ onto a common system. \\
\indent The maximum observed rotation velocity ({\it V}$_{m}$) was measured for each galaxy.  This parameter
was measured by taking the average of the absolute values of the peak velocities of the left and right wings 
of the rotation curve.  The values for each merger remnant are listed in column (6) of Table 3.
\clearpage
{
\begin{deluxetable}{lcccccccc}
\tabletypesize{\normalsize}
\setlength{\tabcolsep}{0.025in}
\tablewidth{0pt}
\tablenum{3}
\pagestyle{empty}
\rotate
\tablecaption{Observed Kinematic and Structural Parameters}
\tablecolumns{9}
\tablehead{
\colhead{Merger} &
\colhead{$\bar{\epsilon}$} &
\colhead{$\bar{\it a_{4}}$} &
\colhead{{\it M}$_{K}$\tablenotemark{a}} &
\colhead{$\sigma$$_{\circ}$\tablenotemark{b}} &
\colhead{{\it V}$_{m}$} &
\colhead{{\it V}$_{m}$/$\sigma$$_{\circ}$} &
\colhead{({\it V}/$\sigma$)$^{*}$} &
\colhead{Template Star\tablenotemark{b}}\\
\colhead{Name} &
\colhead{} &
\colhead{} &
\colhead{} &
\colhead{(mag)}&
\colhead{(km s $^{-1}$)} &
\colhead{(km s $^{-1}$)} &
\colhead{} &
\colhead{star/type}
}
\startdata
UGC 6         &0.19 $\pm$ 0.01  &-0.00036 $\pm$ 0.01245  &-24.01   &220 $\pm$ 10   &60   $\pm$ 9    &0.27  &0.56  &HD 332389 G0III\\ 
NGC 34        &0.11 $\pm$ 0.01  & 0.01361 $\pm$ 0.00882  &-24.61   &201 $\pm$ 8    &100  $\pm$ 4    &0.49  &1.41  &HD 332389 G0III\\
NGC 455       &0.21 $\pm$ 0.01  &-0.00413 $\pm$ 0.00626  &-24.64   &234 $\pm$ 7    &132  $\pm$ 9    &0.56  &1.08  &HD 100059 K0III\\
NGC 1210      &0.08 $\pm$ 0.01  & 0.00246 $\pm$ 0.00284  &-23.72   &247 $\pm$ 6    &34   $\pm$ 12   &0.13  &0.46  &HD 283778 M0III\\
NGC 1614      &0.13 $\pm$ 0.02  & 0.00759 $\pm$ 0.01251  &-24.74   &146 $\pm$ 12   &108  $\pm$ 9    &0.73  &1.91  &HD 332389 G0III\\
AM 0612-373   &0.11 $\pm$ 0.02  &-0.00643 $\pm$ 0.01605  &-25.65   &303 $\pm$ 8    &86   $\pm$ 8    &0.28  &0.80  &HD  99724 K3III\\
NGC 2418      &0.16 $\pm$ 0.01  &-0.00249 $\pm$ 0.00403  &-25.31   &288 $\pm$ 10   &84   $\pm$ 7    &0.29  &0.66  &HD 100347 G8III\\
NGC 2623      &0.24 $\pm$ 0.01  & 0.02390 $\pm$ 0.00956  &-24.22   &191 $\pm$ 7    &40   $\pm$ 4    &0.20  &0.37  &HD 100347 G8III\\
UGC 4635      &0.34 $\pm$ 0.01  & 0.00757 $\pm$ 0.00403  &-24.71   &251 $\pm$ 7    &119  $\pm$ 10   &0.47  &0.65  &HD 100347 G8III\\
NGC 2655      &0.19 $\pm$ 0.01  & 0.00047 $\pm$ 0.00598  &-23.70   &169 $\pm$ 11   &83   $\pm$ 6    &0.49  &1.00  &HD 100347 G8III\\
NGC 2782      &0.26 $\pm$ 0.01  & 0.01028 $\pm$ 0.01193  &-23.83   &196 $\pm$ 8    &99   $\pm$ 7    &0.50  &0.85  &HD 100347 G8III\\
NGC 2914      &0.35 $\pm$ 0.01  & 0.00684 $\pm$ 0.00431  &-23.51   &186 $\pm$ 4    &172  $\pm$ 2    &0.92  &1.26  &HD 100059 K0III\\
UGC 5101      &0.18 $\pm$ 0.01  & 0.00622 $\pm$ 0.01132  &-25.50   &287 $\pm$ 11   &175  $\pm$ 6    &0.60  &1.29  &HD 100347 G8III\\
NGC 3256      &0.19 $\pm$ 0.03  & 0.01348 $\pm$ 0.02181  &-24.72   &241 $\pm$ 16   &49   $\pm$ 7    &0.20  &0.41  &HD 100347 G8III\\
Arp 156       &0.17 $\pm$ 0.01  & 0.01586 $\pm$ 0.01126  &-25.81   &288 $\pm$ 8    &157  $\pm$ 8    &0.54  &1.20  &HD 260158 K0III\\
NGC 3597      &0.40 $\pm$ 0.01  & 0.01113 $\pm$ 0.01406  &-23.72   &174 $\pm$ 9    &140  $\pm$ 9    &0.80  &0.95  &HD 100347 G8III\\
NGC 3921      &0.21 $\pm$ 0.01  & 0.00989 $\pm$ 0.00882  &-25.13   &222 $\pm$ 5    &117  $\pm$ 6    &0.52  &1.02  &HD 100347 G8III\\
NGC 4004      &0.62 $\pm$ 0.01  & 0.02582 $\pm$ 0.02128  &-22.89   &33  $\pm$ 2    &20   $\pm$ 3    &0.60  &0.47  &HD 100347 G8III\\
NGC 4194      &0.24 $\pm$ 0.02  & 0.00914 $\pm$ 0.01868  &-23.21   &116 $\pm$ 7    &81   $\pm$ 4    &0.69  &1.24  &HD 100347 G8III\\
NGC 4441      &0.17 $\pm$ 0.01  & 0.00874 $\pm$ 0.00795  &-22.98   &139 $\pm$ 6    &60   $\pm$ 5    &0.43  &0.94  &HD 100347 G8III\\
AM 1255-430   &0.28 $\pm$ 0.01  &-0.00645 $\pm$ 0.01063  &-24.93   &243 $\pm$ 3    &52   $\pm$ 8    &0.21  &0.33  &HD 99724 K3III\\
NGC 5018      &0.25 $\pm$ 0.01  & 0.01175 $\pm$ 0.00324  &-25.15   &222 $\pm$ 3    &96   $\pm$ 6    &0.43  &0.74  &HD 100059 K0III\\
Arp 193       &0.49 $\pm$ 0.01  & 0.00817 $\pm$ 0.01285  &-24.40   &172 $\pm$ 8    &111  $\pm$ 6    &0.64  &0.65  &HD 332389 G0III\\
AM 1419-263   &0.27 $\pm$ 0.01  & 0.00375 $\pm$ 0.00581  &-24.94   &260 $\pm$ 6    &69   $\pm$ 10   &0.26  &0.43  &HD 100059 K0III\\
UGC 9829      &0.41 $\pm$ 0.01  & 0.02622 $\pm$ 0.00965  &-24.96   &134 $\pm$ 4    &112  $\pm$ 6    &0.83  &1.00  &HD 100059 K0III\\
NGC 6052      &0.44 $\pm$ 0.03  & 0.01377 $\pm$ 0.03115  &-23.55   &80  $\pm$ 5    &43   $\pm$ 4    &0.53  &0.60  &HD 100347 G8III\\
UGC 10607     &0.24 $\pm$ 0.01  & 0.00256 $\pm$ 0.00590  &-25.20   &211 $\pm$ 5    &119  $\pm$ 4    &0.56  &0.99  &HD 100347 G8III\\
UGC 10675     &0.18 $\pm$ 0.02  & 0.00724 $\pm$ 0.01413  &-24.80   &177 $\pm$ 6    &48   $\pm$ 7    &0.27  &0.57  &HD 100347 G8III\\
AM 2038-382   &0.17 $\pm$ 0.01  &-0.00026 $\pm$ 0.00595  &-24.70   &257 $\pm$ 8    &142  $\pm$ 3    &0.55  &1.21  &HD 100347 G8III\\
AM 2055-425   &0.05 $\pm$ 0.02  & 0.00751 $\pm$ 0.01598  &-25.08   &185 $\pm$ 6    &38   $\pm$ 7    &0.20  &0.88  &HD 332389 G0III\\
NGC 7135      &0.18 $\pm$ 0.01  & 0.00393 $\pm$ 0.00474  &-23.95   &277 $\pm$ 9    &108  $\pm$ 3    &0.38  &0.82  &HD 100347 G8III\\
UGC 11905     &0.24 $\pm$ 0.01  &-0.00762 $\pm$ 0.00626  &-24.51   &222 $\pm$ 9    &81   $\pm$ 6    &0.36  &0.64  &HD 100347 G8III\\
NGC 7252      &0.07 $\pm$ 0.01  & 0.00236 $\pm$ 0.00820  &-24.84   &166 $\pm$ 5    &65   $\pm$ 4    &0.39  &1.42  &HD 100347 G8III\\
AM 2246-490   &0.05 $\pm$ 0.02  & 0.00310 $\pm$ 0.01637  &-25.52   &267 $\pm$ 7    &34   $\pm$ 8    &0.12  &0.54  &HD 100347 G8III\\
IC 5298       &0.08 $\pm$ 0.01  & 0.00807 $\pm$ 0.00722  &-24.92   &193 $\pm$ 6    &21   $\pm$ 5    &0.10  &0.36  &HD 100347 G8III\\
NGC 7585      &0.29 $\pm$ 0.01  & 0.00328 $\pm$ 0.00499  &-24.98   &211 $\pm$ 4    &29   $\pm$ 7    &0.13  &0.21  &HD 100059 K0III\\
NGC 7727      &0.24 $\pm$ 0.01  &-0.01634 $\pm$ 0.00936  &-24.23   &231 $\pm$ 5    &154  $\pm$ 5    &0.66  &1.18  &HD 99724 K3III\\ 
\enddata
\tablecomments{All distance dependent values listed in Table 3 assume a value of {\it H}$_{\circ}$ = 75 km s$^{-1}$ Mpc$^{-1}$ and are
derived using only the Heliocentric recessional velocities listed in Table 1.  No additional corrections have been made.
(a) Data from Paper I.  (b)  Data from Paper II.}
\end{deluxetable}

}
\clearpage

\section{Results}
\indent In this section the photometric and spatially resolved kinematic properties of the observed merger remnants listed
in Table 3 are compared with both simulated merger remnants (gas-free and with a gaseous non star-forming component) and a sample of 
84 elliptical galaxies. In \S 5.1 the spatially resolved kinematics of the merger remnants are used to compare
the rotational velocities with random motions to determine whether rotation or anisotropy is the dominant support mechanism.
These results are then compared with both simulations and elliptical galaxies. Section 5.2 probes whether
the merger remnants show the same correlations between kinematic and photometric properties, including shape
and luminosity, as elliptical galaxies.  These correlations are then tested against simulated merger remnants with and without 
a gaseous component.  The simulations are also used to attempt to constrain the progenitor mass ratios of the 
merger remnants.

\subsection{Are Merger Remnants Supported by Rotation or Anisotropy?}
\subsubsection{V$_{m}$/$\sigma$$_{\circ}$}
\indent One method of determining the amount of anisotropic or rotational support in an elliptical galaxy is 
the {\it V}/$\sigma$-$\epsilon$ plane.  This measures the ratio of the maximum observed rotational 
velocity to the velocity dispersion compared with the ellipticity
\markcite{1978MNRAS.183..501B}({Binney} 1978).  Figure 1 is a plot of {\it V}$_{m}$/$\sigma$$_{\circ}$ 
versus $\bar{\epsilon}$.  This plot uses the central velocity dispersion $\sigma$$_{\circ}$ and
mean ellipticity within the radius covered by the kinematic measurements.  The choice in the particular ellipticity and
velocity dispersion used is neccesitated by the need to compare the observed merger remnants 
with the same parameters used in the comparison sample of
elliptical galaxies and numerical models.  Furthermore, \markcite{1981seng.proc...55B}{Binney} (1981), suggests that the
central velocity dispersion is the best measure of the velocity dispersion to compare with the rotational
velocities.  The superposed solid line is a model for an isotropically rotating oblate spheroid
\markcite{1978MNRAS.183..501B}({Binney} 1978).  This line represents objects flattened by rotation as described by the tensor virial theorem.
The equation of the line is:
\begin{equation} 
(\frac{V}{\sigma})_{ISO} \simeq \sqrt{\frac{\epsilon}{(1-\epsilon)}}
\end{equation}
from \markcite{1982modg.proc..113K}{Kormendy} (1982) and is a valid approximation for 0 $<$ $\epsilon$ $<$ 0.95.  The dashed line
is an approximation of the median value of an isotropic rotating {\it prolate} (tumbling end over end) spheroid 
\markcite{1978MNRAS.183..501B}({Binney} 1978).  If the sample objects were prolate, then half the points should lie above the line 
and half should lie below. Plotted in Figure 1 are the merger remnants (filled circles), the 84 galaxies from the comparison 
sample of ellipticals, and the probability contours of the gas-free simulated merger remnants from NB03. 
The over-plotted ellipticals are divided by isophotal shape, open boxes are boxy ellipticals 
and open, flattened diamonds, are disky ellipticals.  The probability contours of the
simulated merger remnants are represented by three shades of gray, the lightest tone represents a 90$\%$ probability of finding
a simulated merger remnant, the middle shade represent a 70$\%$ probability and the darkest shade a 50$\%$ probability.
The figure is divided into four panels, each corresponding to the mass ratios of the simulated merger remnants from NB03.
The mass ratios are noted in each panel.\\ 
{
\begin{figure}
\plotone{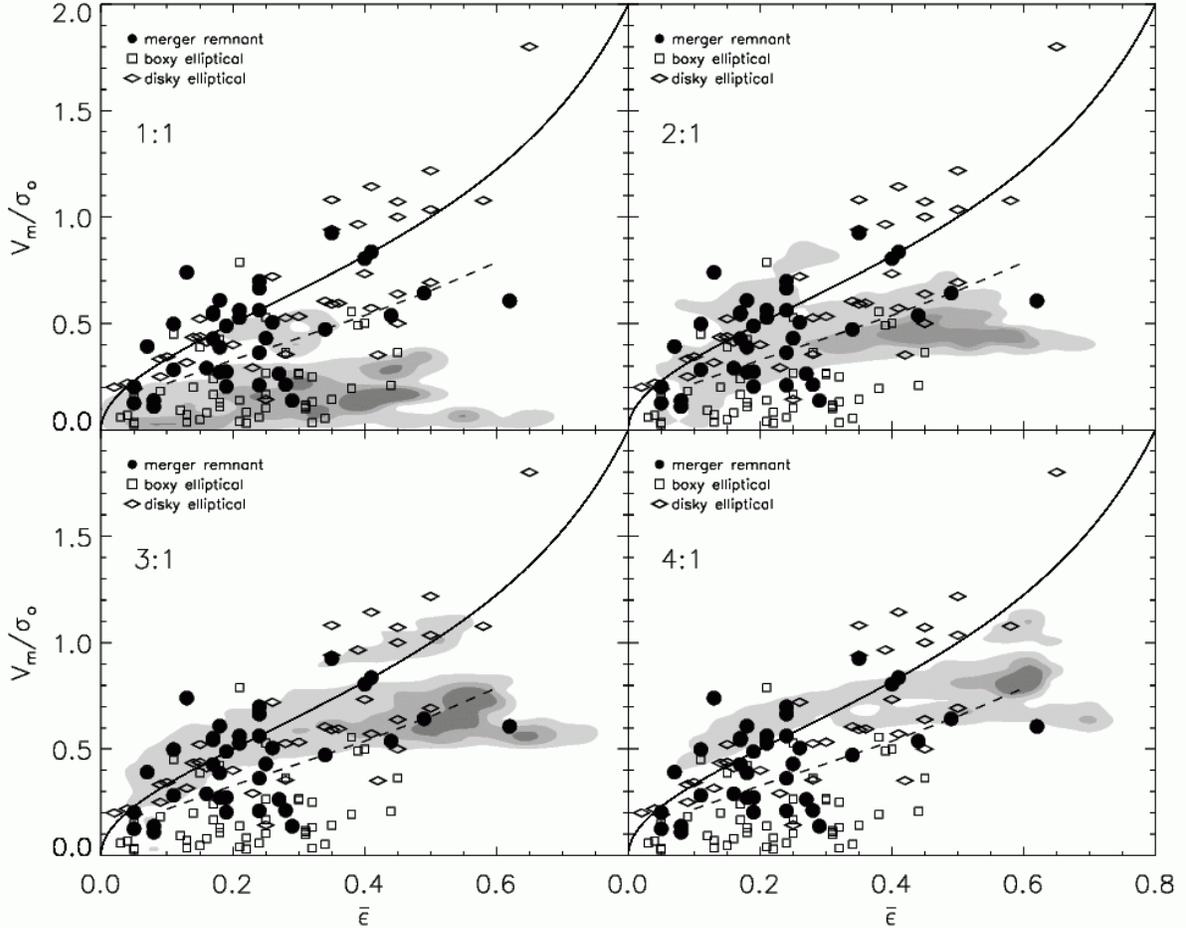}
\caption{Diagram plotting the ratio of the maximum rotational velocity to the central velocity 
dispersion versus the mean ellipticity within the radius of the kinematic measurements.  
Plotted are the merger remnants (filled circles), the 84 galaxies from the comparison sample of ellipticals and the probability
contours of the gas-free simulated merger remnants. The ellipticals are divided by isophotal shape, 
open boxes represent boxy ellipticals and open, flattened diamonds represent disky ellipticals.
The probability contours of the simulated merger remnants are represented by three shades of gray, from light to dark 
representing a 90$\%$, 70$\%$, and 50$\%$ probability respectively of finding a simulated merger remnant.
The solid line is a model of an isotropic rotating oblate spheroid and the dashed line
is an approximation of the median value of an isotropic rotating prolate spheroid \markcite{1978MNRAS.183..501B}({Binney} 1978).  
If the objects are prolate, then half should lie above the line and half should lie below. The figure is divided into 
four panels corresponding to 1:1, 2:1, 3:1 and 4:1 mass ratios of the progenitors of the simulated remnants.}
\end{figure}
}
\indent The majority of merger remnants in Figure 1 appear to cluster close to the line of the isotropic rotating oblate 
spheroid. One-third of the merger remnants lie on or above the line.  The distribution of merger remnants does 
not seem to be consistent with a prolate spheroid, as most of the points lie on or above the 
theoretical median line.  Compared with the elliptical galaxies, the merger remnants 
show a similar range in ellipticities, but not in {\it V}$_{m}$/$\sigma$$_{\circ}$.
There appears to be both an upper and lower limit.  The merger remnants lie in a range of 
0.10 $\leq$ {\it V}$_{m}$/$\sigma$$_{\circ}$ $\leq$ 0.92.  This range is coincidental with only two-thirds
of the ellipticals plotted in Figure 1.  There are 19 ellipticals below the lower limit, all of which 
are boxy in shape and are the brightest ellipticals in the comparison sample.  At the higher limit 
there are 10 ellipticals with {\it V}$_{m}$/$\sigma$$_{\circ}$ $\geq$ 0.92, all of which have disky isophotes.  \\
\indent In order to test whether a substantive difference in {\it V}$_{m}$/$\sigma$$_{\circ}$ values exists 
between the merger remnants and elliptical galaxies, a Kolmogorov-Smirnov (K-S) two-tailed test was employed.
The K-S test probes the null hypothesis that the two distributions in question arise from the same parent population.  
It is a non-parametric test which makes no assumptions about the form of the parent distribution.  
The only assumption is that the two distributions are continuous.  The results indicate that the null hypothesis can
be rejected at better than the 0.02 significance level.  However, while the K-S test is useful for detecting shifts in the
probability distribution, it has difficulty detecting spreads in the distribution.  Such spreads are most noticeable at
the tails of the distribution \markcite{1992nrfa.book.....P}({Press} {et~al.} 1992).  A modified version of the K-S test
developed by Kuiper  is sensitive to changes at the tail ends of the distribution 
\markcite{1962...PKNAvWSerA..63..38,1965...Biometrika..52..309}({Kuiper} 1962; {Stephens} 1965).  The Kuiper test indicates that the null hypothesis can
be rejected at better than the 0.01 significance level.  Thus, the the values of  {\it V}$_{m}$/$\sigma$$_{\circ}$
for the merger remnants and elliptical galaxies do not arise from the same parent population.  An additional item
to keep in mind is the difference in sample sizes between the merger remnants and ellipticals (37 and 84 objects
respectively).  However, a comparison between the critical value, {\it D}, which is used for
the both the K-S and Kuiper tests to determine at what significance level the null hypothesis can or cannot be rejected, produces
only a very small difference between two equal samples of 84 objects and two samples, one of 37 objects and one of 84 objects.\\
\indent This result seems to suggest that the difference in range of {\it V}$_{m}$/$\sigma$$_{\circ}$ values is real.  
It is possible that the lower limit of {\it V}$_{m}$/$\sigma$$_{\circ}$ may provide a means of discriminating between bright 
ellipticals formed in a merger scenario and those possibly formed by slow buildup from accreting nearby galaxies
or multiple mergers.  The upper
limit of {\it V}$_{m}$/$\sigma$$_{\circ}$ may also be a discriminant.  All of these objects are
disky ellipticals with lower luminosity.  This is an interesting limitation
in light of the results in which no {\it single-nuclei} merger remnants, either in this work, or the few
previous observations  (i.e. \markcite{1986ApJ...310..605L}{Lake} \& {Dressler} (1986) and \markcite{2001ApJ...563..527G}{Genzel} {et~al.} (2001)) 
have been observed to have a value of {\it V}$_{m}$/$\sigma$$_{\circ}$ $>$ 0.92. \\
\indent The comparison between the simulated merger remnants and the observed merger remnants produces some surprising
results.  Only 20$\%$ of the observed merger remnants lie within the probability of the 1:1 mass-ratio simulated
merger remnants.  There is a far greater overlap between the 2:1 and 3:1 mass-ratio simulated merger remnants.
This would {\it suggest} that the majority of the observed merger remnants stem from unequal mass mergers 
and are likely to be disky and rotationally supported.  However, an important caveat to keep in mind is that
while the simulated merger remnants are the result of dissipationless merging, the observed merger remnants 
show photometric and spectroscopic evidence of gaseous dissipation.  As noted earlier, there is both numerical 
and observational evidence for the presence of gaseous disks in merger remnants.  \markcite{2000MNRAS.312..859S}{Springel} (2000) 
noted a strong correlation between gaseous dissipation and disky isophotal shapes in his simulated merger remnants.\\
\clearpage
{
\begin{figure}
\plotone{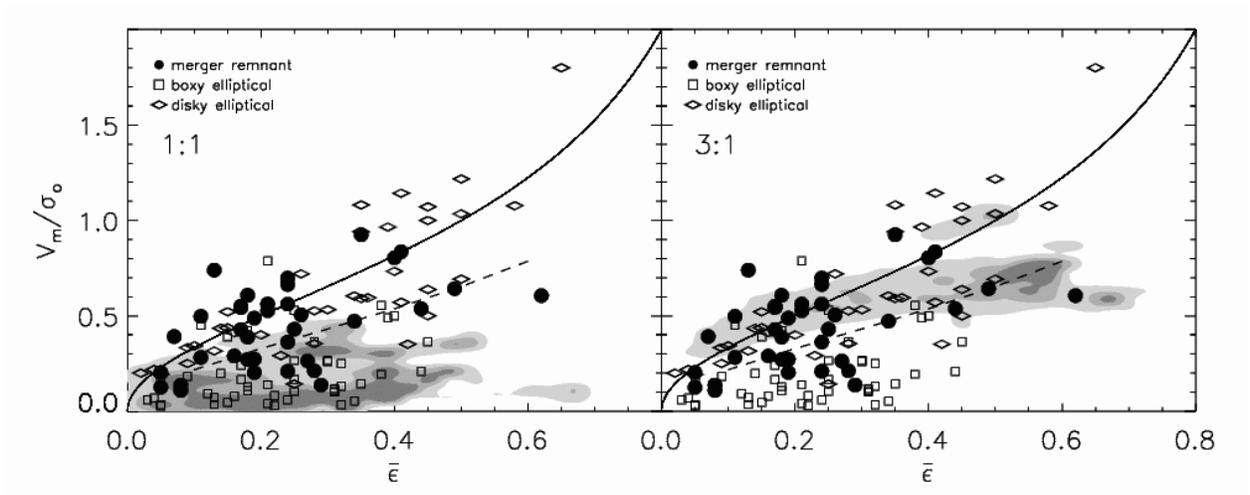}
\caption{{\it V}/$\sigma$-$\epsilon$ plot comparing the merger remnants, elliptical galaxies and the simulated
merger remnants with a gaseous component from N06.  The symbols, overplotted lines and probability
contours have the same meaning as in Figure 1.} 
\end{figure}
}
\clearpage
\indent Figure 2 shows a comparison between the simulated remnants from N06, which include a gas component, and
the observed merger remnants.  The symbols in Figure 2 have the same meaning as in Figure 1.  Only 1:1 and
3:1 mass ratio simulated merger remnants are shown in Figure 2.  The presence of a gaseous component in the simulated
merger remnants does alter their observed parameters. Simulated merger remnants with 1:1 mass ratio progenitors
now span a larger range of {\it V}$_{m}$/$\sigma$$_{\circ}$ values. The contours indicate that more of these simulated remnants
can now be found closer to the line of an oblate isotropic rotator.  The ellipticities lie within a smaller range
(0-0.45) for values of {\it V}$_{m}$/$\sigma$$_{\circ}$ $\leq$ 0.1.  As a result of these differences, more of
the observed merger remnants now lie within the probability countours of the 1:1 mass ratio simulations and fewer
lie within the probability contours of the 3:1 mass ratio simulations.  This suggests that if the mergers contain
a gaseous component, they are more in-line with equal mass ratio progenitors.  These simulations
do not account for the effects of star-formation, which may further alter the observed parameters.  It is possible 
that once a star-formation component is taken into account, nearly all of the observed merger remnants could
be consistent with equal mass ratio progenitors.    

\subsubsection{(V/$\sigma$)$^{*}$}
\indent Another test of rotation versus anisotropy is to compare the measured {\it V}$_{m}$/$\sigma$$_{\circ}$
values with those from the isotropic rotating oblate models (the overplotted black line in Figure 1).
This parameter is known as ({\it V}/{$\sigma$))* and is defined by the following equation:
\begin{equation} 
(\frac{V}{\sigma})^{*}  = \frac{ {\frac{V_{m}}{\sigma_{\circ}}}}{(\frac{V}{\sigma})_{ISO}}
\end{equation}
where ($\frac{V}{\sigma}$)$_{ISO}$ is the approximation noted above and  defined by \markcite{1982modg.proc..113K}{Kormendy} (1982).
A value of ({\it V}/$\sigma$)* = 1.0 defines an object flattened by rotation, values of  
({\it V}/$\sigma$)* $\leq$ 0.7 define anisotropically supported objects, and values between 
0.7 $<$ ({\it V}/$\sigma$)* $<$ 1.0 define objects primarily 
supported by rotation \markcite{1988A&A...193L...7B}({Bender} 1988). ({\it V}/$\sigma$)* has been calculated for
each merger and listed in column (9) of Table 3.\\
\indent The results indicate that just over half of the merger remnants ($\sim$ 57$\%$) appear to be supported primarily by 
rotation.  When compared with the sample of ellipticals, the merger remnants show an apparent lower cutoff.
No merger remnants in the sample have values of ({\it V}/$\sigma$)* $<$ 0.21.  Unlike {\it V}$_{m}$/$\sigma$$_{\circ}$,
there is no apparent cutoff for high values of ({\it V}/$\sigma$)*.  \\
\indent Figure 3 shows a histogram distribution 
comparing the sample of 37 merger remnants with the 84 ellipticals galaxies in the comparison sample.  
The solid line is the distribution of ellipticals, the dashed line represents the merger remnants.
The bin size is 0.1 and the histogram data are plotted to show the fraction of the total sample in each
bin.  The over-plotted down arrows at the top of the histogram represent the peaks of the 1:1, 2:1, 3:1, 
and 4:1 mass ratio dissipationless simulations.  The over-plotted up arrows at the bottom
of the histogram represent the peaks of the 1:1 and 3:1 mass ratio simulations which contain a gaseous component.  The
ratios have been denoted with a $^{\bullet}$ to differentiate them from the gas-free simulations.\\
\indent The distribution shapes do not appear to be similar.  The elliptical galaxies peak at the smallest ({\it V}/$\sigma$)*
values.  The number of ellipticals as a fraction of the total sample decreases as the value of 
({\it V}/$\sigma$)* increases.  The merger remnants show a different distribution shape in ({\it V}/$\sigma$)*.
A K-S was used to test whether the merger remnants and ellipticals arise from the same parent population.
The results indicate that at the 0.02 significance level the hypothesis can be rejected that the ({\it V}/$\sigma$)$^{*}$ distributions 
of the merger remnants and comparison sample of ellipticals arise from the same parent population.  A Kuiper test
was also employed for the reasons noted earler.  The results are somewhat different.  The hypothesis that the two populations
arise from the same parent population can only be rejected at the 0.1 significance level.  \\
\clearpage
{
\begin{figure}
\plotone{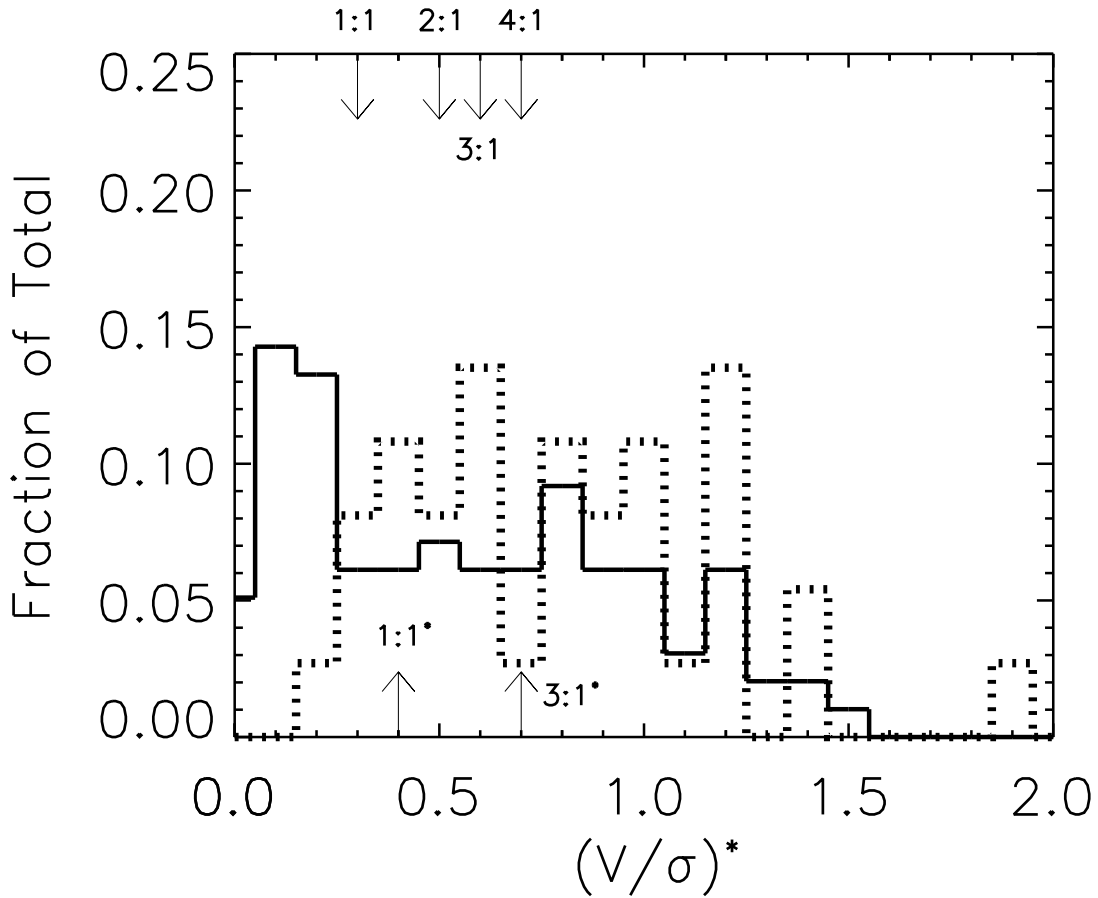}
\caption{Histogram comparing the distributions of ({\it V}/$\sigma$)$^{*}$ between the sample of merger remnants
(dashed line) and the comparison sample of elliptical galaxies (solid line).  The bin size is 0.1 and the
distributions are shown so that each bin is a fraction of the total. The over-plotted down arrows at the top of the histogram
represent the peaks of the 1:1, 2:1, 3:1, and 4:1 mass ratio gas-free simulations.  The over-plotted up arrows at the bottom
of the histogram represent the peaks of the 1:1 and 3:1 mass ratio simulations which contain a gaseous component.  The ratios
have been denoted with a $^{\bullet}$ to differentiate them from the gas-free simulations.}
\end{figure}
}
\clearpage
\indent The overall results of whether merger remnants are rotationally or anisotropically supported show that
a majority of the sample appear to be flattened by rotation.  A significant fraction of the sample do
show evidence of anisotropic support. Furthermore, using both the {\it V}$_{m}$/$\sigma$$_{\circ}$-$\epsilon$
and ({\it V}/$\sigma$)$^{*}$ diagnostics, there appears to be a lower limit to the anisotropy present in
merger remnants.  No merger remnants lie below the values of {\it V}$_{m}$/$\sigma$$_{\circ}$ = 0.10 and 
({\it V}/$\sigma$)$^{*}$ = 0.21.
While it is possible that a much larger sample of merger remnants may include objects with these values, it is also 
possible that this represents a real physical limit.  Thus, these parameters may be a viable method for distinguishing
whether an elliptical galaxy was formed in a single merging event or a slow acrretion-driven buildup, possibly
with multiple mergers, over a large period of time.\\
\indent A comparison with the simulated dissipationless merger remnants from NB03 shows some differences as well.  
The 1:1 mass-ratio simulated merger remnants show a peak at ({\it V}/$\sigma$)$^{*}$ = 0.3, with 80$\%$ having values $\leq$ 0.4.  
The peak of the observed ellipticals lies is 0.1, while the merger remnants seem to have multiple peaks beyond
0.4.  The 2:1, 3:1, and 4:1 mass ratio simulated remnants show peaks at ({\it V}/$\sigma$)$^{*}$ = 0.5, 0.6, and 0.7
respectively.  The simulated remnants with gas show a peak at 0.4 for 1:1 mass ratios and 0.7 for 3:1 mass ratios.
As expected, these values are somewhat larger than the gas-free models.  While the 1:1 mass-ratio simulations with gas
are consistent with one of the peaks of the observed merger remnants, the remainder of the peaks 
do not overlap quite so well.  The elliptical galaxies show a peak at a much smaller value of ({\it V}/$\sigma$)$^{*}$
than either the observed or simulated merger remnants.  The gas-free simulations do not match well with the observed
merger remnants.

\subsection{Correlations Among Kinematics and {\it K}-Band Photometry}
\subsubsection{Rotation and Anisotropy vs. Shape}
\indent It has been well established that elliptical galaxies show a strong correlation between
({\it V}/$\sigma$)$^{*}$ and $\bar{a_{4}/a}$ (e.g. BDM88, B89, \markcite{1996ApJ...464L.119K}{Kormendy} \& {Bender} (1996)).  Anisotropically supported
ellipticals have boxy isophotes, while rotationally supported ellipticals have disky isophotes.  The correlation
is not absolute, there are several examples of elliptical galaxies which run counter to this, however, over a large number
of objects, it is considered to be a well defined trend.  Figure 4 is a plot of the 
({\it V}/$\sigma$)$^{*}$-$\bar{a_{4}/a}$ plane for the merger sample, comparison sample of ellipticals and
the dissipationless simulated merger remnants.  The symbols are the same as in Figure 1. 
The vertical dashed line marks the separation between
boxy (negative) and disky (positive) isophotal shapes.  The horizontal dashed line marks the separation between
anisotropically supported (({\it V}/$\sigma$)$^{*}$ $\leq$ 0.7) and rotationally supported galaxies.
The light, midtone, and dark gray contours represent the 90$\%$, 70$\%$, and 50$\%$ probability  respectively
of finding a simulated merger remnant. The observed trend for ellipticals should place objects in either the bottom 
left quadrant or the top right quadrant.  \\
\indent More than half of the merger remnants follow the expected correlations; 4/37 are boxy and anisotropically 
supported and 17/37 are disky and rotationally supported.  However, the merger remnants do show a somewhat wider dispersion 
in $\bar{a_{4}/a}$ compared with the ellitpical galaxies. In particular, 3 objects appear to be somewhat distant outliers
as compared with both the elliptical galaxies and other merger remnants.  The wider dispersion in $\bar{a_{4}/a}$
may be connected with the results of Paper I, specifically, that these objects have not fully phase mixed.
A significant fraction of the sample do appear to have properties counter to the expected correlations, 
i.e. boxy and rotationally supported or disky and anisotropically supported.  Yet, as can be seen from Figure 4, 
there are elliptical galaxies with similar properties.  A comparison between these elliptical galaxies
and merger remnants does show an interesting result.  Twice as many merger remnants are disky and anisotropically supported
as compared with elliptical galaxies.  This is somewhat unusual, given that there are 2.2 times more elliptical galaxies than
merger remnants.  \\
\clearpage
{
\begin{figure}
\plotone{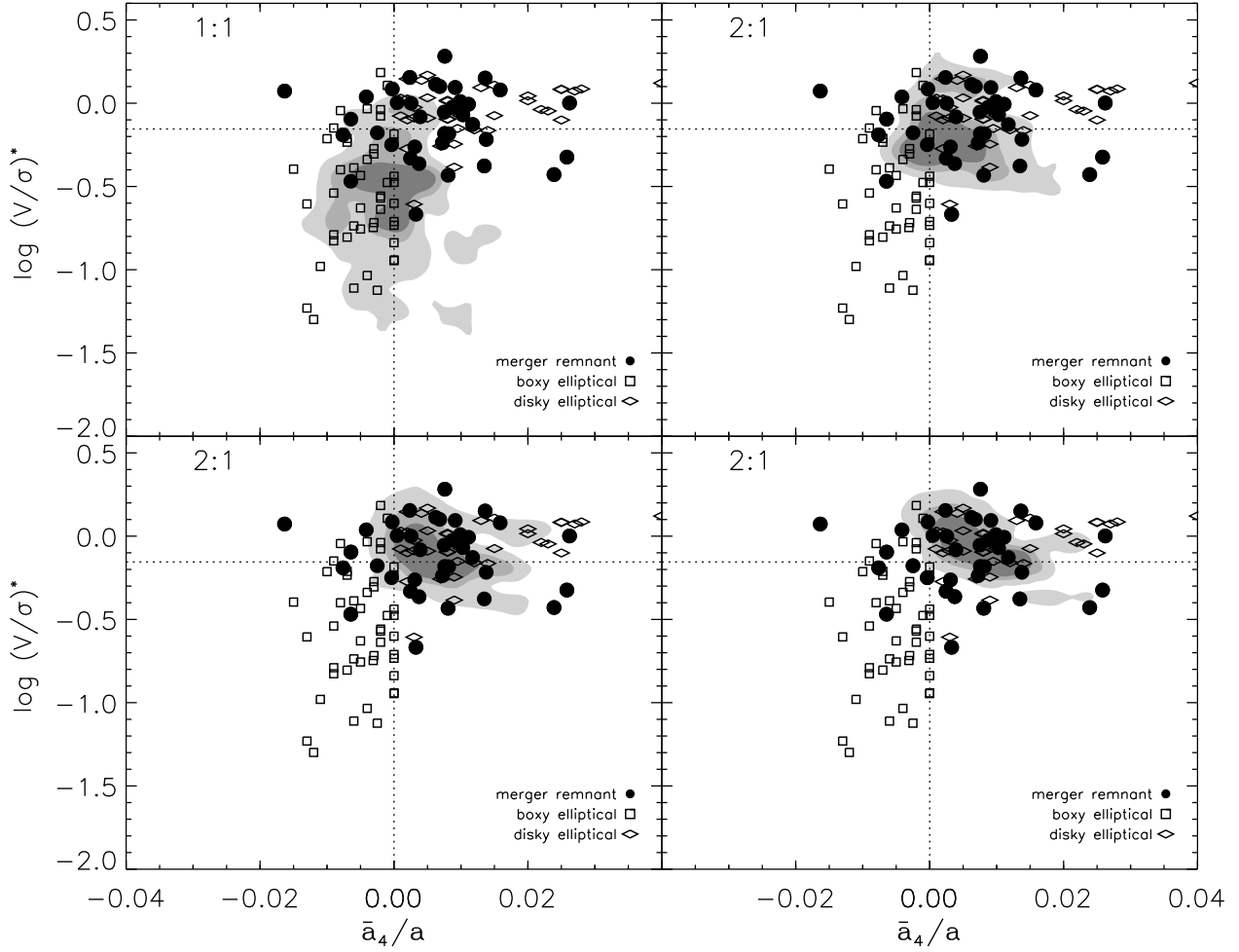}
\caption{Plot showing the plane of isophotal shape versus versus the degree of rotational or anisotropical
support (({\it V}/$\sigma$)$^{*}$ versus $\bar{a_{4}/a}$).  The symbols are the same as in Figure 1. The
light, mid-tone and dark gray contours represent the 90$\%$, 70$\%$, and 50$\%$ probability  respectively
of finding a simulated dissipationless merger remnant from NB03.}
\end{figure}
}
\clearpage
{
\begin{figure}
\plotone{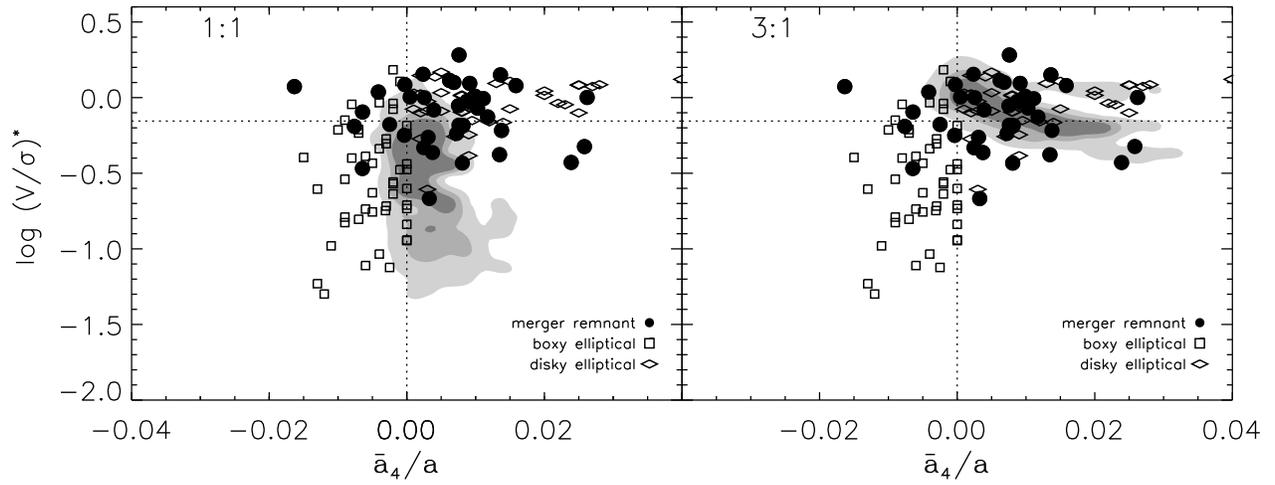}
\caption{Similar plot to Figure 4, however, the contours represent the results from the N06 simulations which
include a gaseous component.}
\end{figure}
}
\clearpage
\indent When compared with the dissipationless simulated merger remnants, approximately one-third of the observed merger remnants 
overlap with the 1:1 mass-ratio simulated remnants.  The 2:1 and 3:1 mass-ratio simulated remnants show a greater overlap with the 
observed merger remnants.  This  seems to suggest that the observed merger remnants are predominantly unequal mass
in nature.  However, NB03 do note an odd result in their models.  They find that 28$\%$ of the 1:1 remnants appear to be 
disky, anisotropic system with the following properties:  $\bar{a_{4}/a}$ $>$ 0.003 and ({\it V}/$\sigma$)$^{*}$ $<$ 0.5.
Simulated merger remnants with these properties are defined as ``forbidden''  because they fail to
resemble observed elliptical galaxies.  Yet, 6/37 merger remnants have these observed properties.  \\
\indent While it is unlikely that the kinematics of the merger remnants will change once a single nucleus is formed 
\markcite{1988ApJ...331..699B,1992ApJ...393..484B,1999Ap&SS.266..195M}({Barnes} 1988, 1992; {Mihos} 1999), the results of Paper I 
indicate that phase mixing for the merger remnants
is far from complete.  This means that there is still some ``memory'' of the progenitor orbits remaining.
In terms of the observed merger remnants, it is possible that these apparently contradictory properties are induced
by gaseous dissipation which can modify isophotal shapes \markcite{1996ApJ...471..115B}{Barnes} \& {Hernquist} (1996).  Once star-formation 
ceases and the merger becomes fully phase-mixed, the isophotal shapes may end up more in line with shapes
expected from their kinematic properties.\\
\indent Figure 5 is a similar plot to Figure 4, however, the simulated merger remnants shown are from N06.  The symbols
have the same meaning as in Figure 1 and only the 1:1 and 3:1 mass ratio simulated remnants are plotted.  The most noticeable
difference between the gas-free simulations and the ones with a gaseous component can be seen in the 1:1 mass ratio
plot (Figure 5 {\it left}).  The probability contours have shifted to the right. so that there is far less coverage where
$\bar{a_{4}/a}$ $\leq$ 0 and more simulated remnants with values of $\bar{a_{4}/a}$ $>$ 0. The ({\it V}/$\sigma$)$^{*}$ values
appear to be less affected by the gaseous component, although the contours now extend to slightly higher values.
The inclusion of a gaseous component seems to produce remnants from equal mass mergers which, although containing 
signficant anisotropy, are quite disky
in shape.  This shift is better able to account for the observed merger remnants, especially those with so-called
``forbidden'' parameters (as noted above).  The net result, however, is that far fewer of the 84 elliptical galaxies
in the comparison sample
now coincide with 1:1 mass ratio progenitors with gaseous components.  The 3:1 mass ratio progenitors with a gaseous
component now extend further into the disky $\bar{a_{4}/a}$ regime.  However, the ({\it V}/$\sigma$)$^{*}$ values
do appear to be affected.  At smaller positive $\bar{a_{4}/a}$ values ($<$ 0.05), the probability
contours barely lie below  ({\it V}/$\sigma$)$^{*}$ $<$ 0.7.  The net result for the observed merger remnants, is
that objects which had been consistent with dissipationless 3:1 mass ratio progenitors are now consistent
with 1:1 mass ratio progenitors {\it with} a gaseous component.  Moreover, between both the 1:1 and 3:1 mass ratio
simulations with gas, practically all of the observed merger remnants with  $\bar{a_{4}/a}$ $>$ 0 lie within one
of the these probability contours.  However, there are now several more mergers with $\bar{a_{4}/a}$ $<$ 0
which are not consistent with any of the simulations.

\subsubsection{Rotation and Anisotropy vs. Luminosity}
\indent \markcite{1977ApJ...218L..43I}{Illingworth} (1977) and \markcite{1983ApJ...266...41D}{Davies} {et~al.} (1983) found that a relationship exists between {\it M}$_{B}$
and both {\it V}$_{m}$/$\bar{\sigma}$ and ({\it V}/$\sigma$)$^{*}$.  Faint ellipticals rotate more rapidly than bright 
ellipticals.  However, \markcite{1983ApJ...266...41D}{Davies} {et~al.} (1983) noted that there was a larger scatter for bright ellipticals, 
possibly due to a combination of varying degrees of anisotropy and projection effects.  \markcite{1989A&A...217...35B}{Bender} {et~al.} (1989)
noted a strong correlation between the {\it B}-Band luminosity and $\bar{a_{4}/a}$.  \markcite{1983ApJ...266...41D}{Davies} {et~al.} (1983)
predicted that if mergers form elliptical galaxies, then {\it V}$_{m}$/$\bar{\sigma}$ should
decrease with increasing mass.  
\clearpage
{
\begin{figure}
\plotone{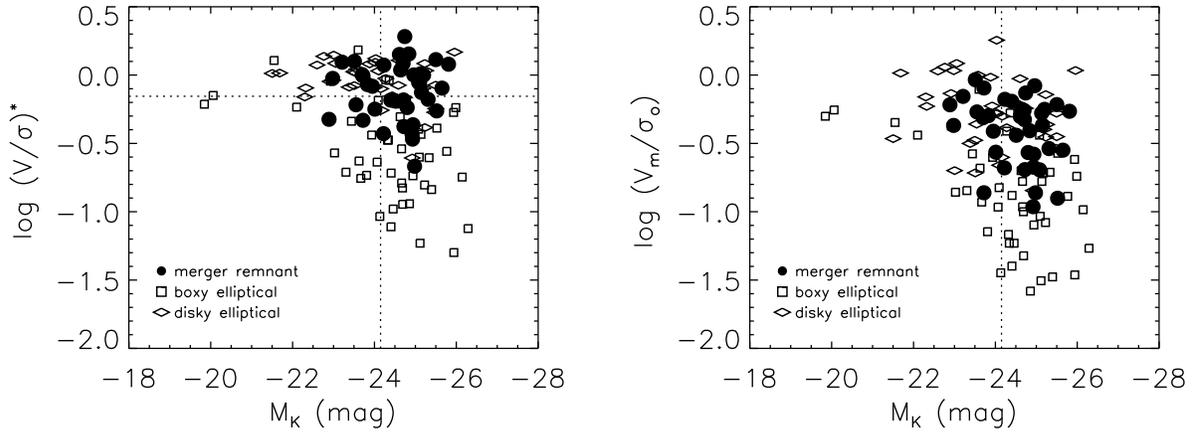}
\caption{(a) {\it (left)} Plot of {\it K}-band absolute magnitude versus degree of anisotropy.  Symbols are the
same as in Figure 1.  The {\it K}-band magnitudes for the ellipticals are taken from 2MASS.
The horizontal dashed line at ({\it V}/$\sigma$)$^{*}$ = 0.7
marks the transition between anisotropically and rotationally supported galaxies.  The vertical dashed line
at {\it M}$_{K}$ = -24.15 indicates the luminosity of a 1{\it L}* elliptical galaxy \markcite{2001ApJ...560..566K}({Kochanek} {et~al.} 2001).
(b) {\it (right)} Plot of {\it K}-band absolute magnitude versus {\it V}$_{m}$/$\sigma$$_{\circ}$.  
Symbols are the same as in Figure 1.  Again, the vertical dashed line
at {\it M}$_{K}$ = -24.15 indicates the luminosity of a 1{\it L}* elliptical galaxy.}
\end{figure}
}
\clearpage
\indent Figures 6a and 6b show the {\it M}$_{K}$-log ({\it V}/$\sigma$)$^{*}$
and {\it M}$_{K}$-log {\it V}$_{m}$/$\sigma$$_{\circ}$ planes respectively for the merger sample and 
the 84 objects in the comparison sample of elliptical galaxies.  The symbols are the same as in Figure 1.
In Figure 6a, the horizontal dashed line at ({\it V}/$\sigma$)$^{*}$ = 0.7 marks the transition between anisotropically 
and rotationally supported galaxies.  In both figures,  The vertical dashed line
at {\it M}$_{K}$ = -24.15 indicates the luminosity of a 1{\it L}* elliptical galaxy \markcite{2001ApJ...560..566K}({Kochanek} {et~al.} 2001), corrected
to {\it H}$_{\circ}$ = 75 km s$^{-1}$ Mpc$^{-1}$.
A Spearman rank correlation test indicates that the elliptical galaxies show a strong anti-correlation
(as luminosity increases both {\it V}$_{m}$/$\sigma$$_{\circ}$ and ({\it V}/$\sigma$)$^{*}$ decrease)
at better than a 0.001 significance level for both ({\it V}/$\sigma$)$^{*}$ and {\it M}$_{K}$ 
and {\it V}$_{m}$/$\sigma$$_{\circ}$ and {\it M}$_{K}$.  The merger remnants show no correlation (or anti-correlation)
between {\it M}$_{K}$ and log ({\it V}/$\sigma$)$^{*}$ and a slight anti-correlation between
{\it M}$_{K}$ and log {\it V}$_{m}$/$\sigma$$_{\circ}$ at the 0.1 significance level. This seems
to weakly follow the prediction of \markcite{1983ApJ...266...41D}{Davies} {et~al.} (1983) that luminosity (and therefore mass) increases
as objects become more anisotropic.

\section{Summary and Discussion}
\indent The main results of this work are as follows:\\
\indent 1. More than half ($\sim$ 57 $\%$) of the merger remnants exhibit properties consistent with the paradigm of
boxy and anisotropic or disky and isotropic rotators:  4/37 are boxy and anisotropically supported; 
17/37 are disky, and rotationally supported.  However,
the correlation with luminosity is less clear.  Most (10/17) of the disky and rotationally supported merger remnants
have luminosities $\ge$ 1{\it L}*.  Yet, statistically, Figure 6b does indicate a weak anti-correlation between
{\it K}-band luminosity and {\it V}$_{m}$/$\sigma$$_{\circ}$, suggesting that as mass increases the merger remnants
are supported more by anisotropy.  Figures 6a and 6b also show that compared with the elliptical galaxies,
a higher fraction of the merger remnants have {\it K}-band luminosities $\ge$ 1{\it L}*.\\
\indent 2.  Kinematically, the sample is split almost evenly between anisotropically supported (16/37)
and rotationally supported (21/37) merger remnants.  The discrepancy between the photometric and kinematic results
may be due, in part, to the possibility that the isophotal shapes are still ``in flux.''  This is consistent
with the findings in Paper I, which showed that the most of the merger remnants in the sample have not fully phase-mixed.
The kinematic properties may be close to, or at, their ``final'' values.  This is supported by numerical simulations
\markcite{1988ApJ...331..699B,1992ApJ...393..484B,1999Ap&SS.266..195M}({Barnes} 1988, 1992; {Mihos} 1999)
which have shown that once a merger has coalesced
to form a single nucleus, kinematic properties such as the velocity dispersion remain the same from that point onwards.\\
\indent 3.  A significant fraction of the sample show trends counter to the expected correlation of boxy and anisotropic or 
disky and rotationally supported.  The merger remnants show a larger scatter in $\bar{a_{4}/a}$ compared with
the elliptical galaxies.  Three merger remnants appear to be outliers compared with the elliptical galaxy sample.
An interesting result from the comparison with NB03, is that 1:1 mass ratio progenitors with
so-called ``forbidden'' properties ($\bar{a_{4}/a}$ $>$ 0.003 and ({\it V}/$\sigma$)$^{*}$ $<$ 0.5) and which are not
observed in elliptical galaxies, are found to exist in the observed merger remnant sample.  28$\%$ of the 1:1 mass ratio
simulated merger remnants in NB03 lie in this ``forbidden regime,'' as do 6/37 ($\sim$ 16$\%$) observed merger
remnants\\
\indent 4.  The results suggest that the amount of anisotropy or pressure support in recent merger remnants is far less
than in some old elliptical galaxies.  There appears to be a clear line 
at {\it V}$_{m}$/$\sigma$$_{\circ}$ = 0.10 and ({\it V}/$\sigma$)$^{*}$ = 0.21 below which no merger remnants are found.  
This may be an observational tool which can be used to discern which ellipticals have been formed from 
disk-disk mergers.\\
\indent 5.  When compared with {\it dissipationless} merger remnants, approximately one-third of the merger remnants
are consistent with 1:1 mass progenitors, the rest are comparable, primarily, with 2:1 and 3:1 mass
progenitors.\\
\indent 6.  When the observed merger remnants are compared with the simulations that include
a gaseous component, the fraction consistent with 1:1 mass ratio progenitors increases
significantly.  The simulations with a gaseous component appear consistent with remnants
that have so-called ``forbidden'' parameters.  The inclusion of a gaseous component appears to be more capable of 
accounting for observed merger remnants which are anisotropically supported, but have disky isophotal shapes.
However, a consequence of this is that fewer boxy, anisotropically supported ellipticals
are consistent with the newer simulations.\\
\indent In the context of the Merger Hypothesis, a fundamental question to raise is: {\it How do merger remnants
relate to the apparent dichotomy in elliptical galaxy types?}  The photometry
and kinematic results from Papers I and II support the idea that mergers are capable of forming elliptical
galaxies.  The data presented here appear to show that mergers can produce both rotationally and anisotropically 
supported systems.  Papers I and II also suggest that most, if not all of the merger remnants have undergone some form of gaseous 
dissipation.  Numerical simulations suggest that the effects of dissipation may contribute primarily
to the formation of {\it disky} ellipticals \markcite{1991ApJ...370L..65B,1996ApJ...471..115B,2000MNRAS.312..859S}({Barnes} \& {Hernquist} 1991, 1996; {Springel} 2000).
In particular, the effects of dissipation can alter the stellar orbits, limiting or suppressing box orbits.
This produces tube orbits which can lead to the formation of disky isophotal shapes.  \markcite{2000MNRAS.312..859S}{Springel} (2000)
modeled the effects of star-formation and feedback for use in hydrodynamical simulations of galaxy formation.
The models cooled gas radiatively with a Schmidt-like star-formation rate and included the effects of feedback
by supernovae.  When applied to mergers of gaseous disk systems, the models produced a strong central starburst.
The process of gaseous dissipation appeared to correlate with the presence of disky isophotes.  
Springel also included dissipationless disk-disk merger merger remnants as a comparison.  Those produced predominantly boxy 
isophotal shapes.  However, like the results of NB03, viewing angle could make equal mass mergers appear either disky or boxy.\\
\indent Moreover, \markcite{2002MNRAS.333..481B}{Barnes} (2002) found that dissipation builds rotating 
gaseous disks. Such gaseous disks have been confirmed observationally (e.g. \markcite{1992ApJ...396..510W,1993AJ....106.1354W}{Wang} {et~al.} (1992); {Whitmore} {et~al.} (1993))
in some mergers.  If these disks can form stars, it may explain both the disky isophotal shapes, as well
as the apparent rotation observed in many of the merger remnants.  These factors may contribute to the possible
``lower-limit'' in ({\it V}/$\sigma$)$^{*}$ observed in the merger remnants.\\  
\indent It has been previously suggested that mergers are incapable of forming 
giant elliptical galaxies (e.g. \markcite{1998ApJ...497..163S,1999MNRAS.309..585J,2001ApJ...563..546C,2001ApJ...563..527G}{Shier} \& {Fischer} (1998); {James} {et~al.} (1999); {Colina} {et~al.} (2001); {Genzel} {et~al.} (2001)).
The photometry and kinematics presented in Papers I and II clearly challenge that assertion, as both the luminosity
and kinematics of a significant number of the observed merger remnants are consistent with giant ellipticals.  However,
the largest measured $\sigma$$_{\circ}$ in the sample is 303 km s$^{-1}$, while observations of elliptical galaxies show objects with
$\sigma$$_{\circ}$ $>$ 350 km s$^{-1}$.  These objects are almost exclusively bright and boxy in shape,
with very small values of ({\it V}/$\sigma$)$^{*}$.  Even the dissipationless 1:1 mass ratio simulations of NB03 do not
appear to be consistent with the {\it most} massive giant ellipticals, and the N06 simulations with gas show even less overlap.  
All of these factors may point to some type of upper limit for the mass of an elliptical galaxy formed from 
disk-disk mergers.  Other mechanisms, such as dissipationless mergers of bulge-dominated galaxies \markcite{2006ApJ...636L..81N}({Naab}, {Khochfar}, \&  {Burkert} 2006) may
be able to better reproduce the elliptical galaxies with the smallest ({\it V}/$\sigma$)$^{*}$ and most negative $\bar{a_{4}/a}$
parameters.  However, this does raise a host
of other problems, including reproducing the Fundamental Plane and M$_{BH}$-$\sigma$$_{\circ}$ relations
(e.g. \markcite{2003MNRAS.342..501N,2005ApJ...623L..67K}{Nipoti}, {Londrillo}, \&  {Ciotti} (2003); {Kazantzidis} {et~al.} (2005)), as well as the presence of intermediate-age stellar populations 
and globular clusters in elliptical galaxies (e.g. \markcite{1990ApJ...364L..33S,1997AJ....114.1797W,2001A&A...379..781R,2002AJ....124..147W,2004MNRAS.351L..19T,2004ApJ...613L.121G,2005MNRAS.362....2T}{Schweizer} {et~al.} (1990); {Whitmore} {et~al.} (1997); {Rejkuba} {et~al.} (2001); {Whitmore} {et~al.} (2002); {Thomas}, {Maraston}, \&  {Korn} (2004); {Goudfrooij} {et~al.} (2004); {Trager} {et~al.} (2005)).  Moreover, with recent deep surveys indicating the
presence a red-sequence in color-magnitude diagrams of galaxies as far back as z $\sim$ 1
\markcite{2004ApJ...601L..29H,2004ApJ...608..752B,2005ApJ...632..191M}({Hogg} {et~al.} 2004; {Bell} {et~al.} 2004; {McIntosh} {et~al.} 2005), new questions have been raised concerning
the limitations of disk-disk mergers versus other mechanisms, including dissipationless mergers between bulge-dominated
galaxies.  Future studies need to address the limitations of {\it both} merger scenarios to help constrain
the importance and relative contribution of each to the formation of present day elliptical galaxies. \\
\indent Another intriguing result, is that a significant fraction of the sample appear to show properties 
{\it contrary} to those expected of the boxy vs. disky elliptical paradigm.  That is, a significant fraction show evidence
of anisotropic kinematics, yet disky isophotal shapes, conversely, there are a few objects which show
evidence of strong rotation, yet have boxy isophotes.  The dissipationless merger models of NB03 also
show similar results.  They note that 28$\%$ of the 1:1 remnants show unexpected properties, in particular,
small ({\it V}/$\sigma$)$^{*}$ values with very disky isophotes.  Further, they remark that
the lack of {\it observed} merger remnants or elliptical galaxies with these properties may constitute a serious problem 
for the Merger Hypothesis.  The results presented in this paper may indicate a reprieve for that possibility,
as some real merger remnants clearly show these properties.  Moreover, the newer simulations of N06,
which include a gaseous component, appear to be more consistent with observed merger remnants which 
are disky and anisotropic.  This raises several questions, such as why there are observed merger remnants as well as simulations
with these properties, but relatively few or no elliptical galaxies.  Are those types of elliptical galaxies
still awaiting discovery?  Or will the observed merger remnants evolve into elliptical galaxies with properties
similar to those in present day ellipticals?  How does
the presence of star-formation and the evolution of the stellar populations fit into this situation?  Perhaps 
future numerical simulations which can take into account both star-formation in the gaseous component and the evolution of the 
stellar populations can provide an answer.  \\
\indent Finally, a comparison between the simulated merger remnants from NB03 and N06 with the observed merger remnants
presented here suggest that most of the sample are consistent either with dissipationless unequal mass mergers
or equal mass mergers with a gasoues component.  The latter is a more realistic result, given the supporting
evidence from Papers I and II, and direct observations (e.g. \markcite{1988ApJ...324L..55S,1990A&A...228L...5D,
1991ApJ...368..112W,1991A&A...251....1C,1992ApJ...396..510W,1994AJ....107...67H,2001ApJ...550..104Y,2004ApJ...616L..67W}{Sanders} {et~al.} (1988); {Dupraz} {et~al.} (1990); {Wang}, {Scoville}, \&  {Sanders} (1991); {Casoli} {et~al.} (1991); {Wang} {et~al.} (1992); {Hibbard} {et~al.} (1994); {Yun} \& {Hibbard} (2001); {Wang} {et~al.} (2004)).\\
\indent The results presented here would benefit from follow-up work.  Most importantly,
kinematic observations along the {\it minor} axis of the merger remnants would produce a stronger
argument for or against rotational or anisotropic support.  It would also help determine if any of the merger remnants are
triaxial, which may explain some of the unexpected results.  Additional kinematic observations out to 
larger radii along the major axis would help determine whether the observed 
{\it V}$_{m}$/$\sigma$$_{\circ}$ and ({\it V}/$\sigma$)$^{*}$ values are upper or lower limits.  Finally,
a more thorough analysis of both optical and infrared spectra can help confirm whether gaseous dissipation has occurred
and the amount of star-formation present.  These results can be used to constrain the evolution of the stellar populations
in merger remnants, which, in conjunction with the kinematic results, can help address the question of whether
these objects are capable of forming the most luminous and most massive elliptical galaxies.

\acknowledgments
We would like to thank Andreas Burkert and Thorsten Naab for providing data from their numerical
models of dissipationless merger remnants as well as the data used to make comparisons with observed elliptical
galaxies.  A special thanks in particular is given to Thorsten Naab for sharing with us his latest unpublished results.
These additional datasets have helped to improve the quality
of the work presented here. We would also like to thank Josh Barnes for informative discussions during the
early stages of preparing this manuscript.  A special thanks is given to Michael Cushing for his help and useful 
suggestions in developing the early versions of the IDL code used to derive the velocity dispersions from the 
extracted spectra. We thank Michael Connelley for taking {\it K}-band data of one of the mergers in the sample and
Amy Apodaca for providing insight into some statistical issues, and finally, as always, a thanks to
the back and forth nature provided by T. B. Wall.
We would also like to thank the anonymous referee for comments and suggestions which have improved the manuscript.
This research has made use of the NASA/IPAC Extragalactic Database
(NED) which is operated by the Jet Propulsion Laboratory, California Institute of Technology,
under contract with the National Aeronautics and Space Administration.  
This publication makes use of data products from the Two Micron All Sky Survey, which is a joint project 
of the University of Massachusetts and the Infrared Processing and Analysis Center, California Institute 
of Technology, funded by the National Aeronautics and Space Administration and the NSF.  This research is supported
in part, by a Fellowship from the NASA Graduate Student Researchers Program, Grant $\#$ NGT5-50396.


\end{document}